\documentclass[aps,pra,twocolumn]{revtex4-2}  % Choose 'preprint' or 'twocolumn'
\usepackage{amsmath,amssymb}  % For mathematical symbols and equations
\usepackage{graphicx}         % For including graphics
\usepackage{booktabs}
\usepackage{natbib}
\usepackage[colorlinks=true, linkcolor=blue, citecolor=blue, urlcolor=blue]{hyperref} % Enable and customize hyperlinks
\usepackage{tabularx}
\usepackage{bm}
\usepackage{ulem}

\def\HIM{Helmholtz Institute Mainz, 55099 Mainz, Germany}
\def\GSI{GSI Helmholtzzentrum für Schwerionenforschung GmbH, 64291 Darmstadt, Germany}
\def\JGU{Johannes Gutenberg University, Mainz 55128, Germany}

\def\FMUV{Faculty of Mathematics, University of Vienna, Oskar-Morgenstern-Platz 1, 1090 Vienna, Austria}
\def\GPGUV{Gravitational Physics Group, University of Vienna, W\"{a}hringer Stra{\ss}e 17, 1090 Vienna, Austria}

\def\UC{Department of Physics, University of California at Berkeley, Berkeley, California 94720-7300, USA}

\begin{document}

\title{Improved constraints on exotic interactions between electron and proton in hydrogen}

\author{Lei Cong$^{1,2,3}$, Filip Ficek$^{4,5}$, Pavel Fadeev$^{3}$, and Dmitry Budker$^{1,2,3,6,*}$}
\address{
$^{1}$ \HIM\\
$^{2}$ \GSI\\
$^{3}$ \JGU\\
$^{4}$ \FMUV\\
$^{5}$ \GPGUV\\
$^{6}$ \UC\\
* budker@uni-mainz.de
 }

\begin{abstract}
Atomic spectroscopy can be used to search for new bosons that carry exotic forces between elementary fermions. A comparison of a recent precise measurement [Bullis \textit{et al.}, Phys. Rev. Lett. \textbf{130}, 203001 (2023)] 
%\cite{bullis_ramsey_2023} 
of the hyperfine splitting of the 2S$_{1/2}$ electronic levels of hydrogen and up-to-date bound-state quantum electrodynamics theory yields improved constraints on electron-proton exotic interactions of the dimensionless coupling strengths $g_Ag_A$, $g_pg_p$, and $g_Vg_V$ corresponding to the exchange of a axial-vector, pseudoscalar (axionlike) or vector boson, respectively.
\end{abstract}

\maketitle

%\section{Introduction}

Precise experiments with atoms and molecules \cite{safronova_search_2018}, color centers in diamond \cite{jiao_experimental_2021}, comagnetometers \cite{ji_constraints_2023,xu_constraints_2025}, and torsion pendulums \cite{heckel_limits_2013}, among others, are successfully employed for exploring potential extensions to the Standard Model (SM), including the search for new particles. Various simple atomic systems \cite{karshenboim_hyperfine_2011,ficek_constraints_2017,ficek_constraints_2018,fadeev_pseudovector_2022}, such as positronium, muonium, helium, antiprotonic helium, antihydrogen and hydrogen, have been utilized to probe spin-dependent exotic interactions \cite{moody_new_1984, dobrescu_spin-dependent_2006, fadeev_revisiting_2019, cong_spin-dependent_2024,cong_searching_2025}.
Recently, a precision record \cite{bullis_ramsey_2023} was set by a measurement of the hyperfine splitting of the 2S$_{1/2}$ electronic levels of hydrogen. This breakthrough offers an opportunity to significantly improve upon previous constraints \cite{karshenboim_hyperfine_2011,fadeev_pseudovector_2022}. 

In this work, we make two advances. First, we significantly improve the constraints on electron–proton exotic interactions.
By comparing experimental data with up-to-date theoretical values from bound-state quantum electrodynamics, %particularly for the relevant quantity $D_{21}$, %\LC{improved}
%\FF{($D_{21}$ is not introduced at this moment, so maybe it's better to write explicitly "hyperfine splitting" instead of "relevant quantity $D_{21}$". Alternatively, I think one could just remove the whole phrase "particularly for the relevant quantity $D_{21}$)}
constraints on electron-proton exotic interactions can be obtained.
%Second, we study a previously unconstrained type of exotic interaction that can be studied via the hydrogen system—the vector–vector type—which, although previously largely overlooked (for example, constraints on $V_2$ and $V_3$ have not been interpreted in terms of vector-vector interaction), merits equal attention based on the latest comprehensive theoretical framework summarized in Ref.\,\cite{cong_spin-dependent_2024}. 
Second, we study a previously unconstrained type of exotic interaction—the vector–vector type—that can be probed via the hydrogen system. Although this interaction has largely been overlooked (for instance, existing constraints on $V_2$ and $V_3$ have not been interpreted in terms of vector–vector interactions), it merits equal attention based on the comprehensive theoretical framework recently summarized in Ref.\,\cite{cong_spin-dependent_2024}.

%According to SM theory, where only conventional electromagnetic interactions are considered, $D_{21}$ is expected to be zero. This expectation arises because hyperfine splitting scales as $1/n^3$. \DB{Are there any corrections to this, for example, relativistic?} 
%The interest in $D_{21}$ was inspired by the experiment \cite{heberle_hyperfine_1956}. 
The hyperfine splitting difference, $D_{21} =  8\Delta_{\text{hfs}}(2s) -\Delta_{\text{hfs}}(1s)$, is a critical parameter for probing the presence of exotic forces in atomic physics. 
In conventional atoms, the sensitivity of tests based on the Hyperfine Structure (hfs) is limited by insufficient knowledge of their nuclear structure \cite{i_eides_theory_2001}. %\FF{
However, since the lowest-order nuclear corrections to $\Delta_{\text{hfs}}(2s)$ and $\Delta_{\text{hfs}}(1s)$ scale with the electron density in the ratio ${1}:{8}$, they do not contribute to the difference $D_{21}$.
Of course some residual contributions to $D_{21}$ from nuclear structure and other standard-model effects \cite{karshenboim_hyperfine_2002,karshenboim_hyperfine_2002-1,yerokhin_electron_2008} still apply.
%}
This provides a chance to test the QED theory of bound states at a level of precision orders of magnitude better than that achievable for the ground-state hyperfine interval $\Delta_{\text{hfs}}(1s)$ alone \cite{karshenboim_precision_2005} and also offers a good opportunity for new physics searches.

%A nonzero value of $D_{21}$ would indicate a deviation from this expected scaling, potentially pointing to the influence of additional, non-standard interactions. Such deviations could arise from exotic spin-spin forces that decay over distance according to a different law, such as $1/r $, instead of the conventional $1/r^3$ dependence observed in dipole-dipole interactions. Measuring $D_{21}$ and observing a non-zero value thus provides a sensitive indicator for new physics.
%\LC{\cite{kolachevsky_measurement_2009} mentioned that $D_{21}$ have a theoretical expectation value 48 kHz. Need to think about this.}

%\LC{What about study $D_{21}$ of deuterium atom and helium ion $^3$He$^+$? there are already data in \cite{karshenboim_precision_2005}. }\LC{experiment data not updated for a long time.}

%Table\,\ref{table1} summarizes the latest theoretical and experimental results as well as the difference ($\mu$) between theory and experiment, and the combined uncertainty ($\sigma$). We use these to derive $\Delta D$, representing the maximum deviation. 
%\FF{
Table \ref{table1} summarizes the current data concerning the value of $D_{21}$, in particular the results of theoretical calculations and experimental measurements, their difference $\mu$, and the combined uncertainty $\sigma= \sqrt{{\sigma_{th}}^2 + {\sigma_{expt}}^2}$. We use these to derive $\Delta D$, representing the maximal deviation at $90\%$ confidence level,
%} 
%The calculations are based on the following equations: $\mu = \text{Theory} - \text{Expt}$\,, $\sigma = \sqrt{{\sigma_{th}}^2 + {\sigma_{expt}}^2}$\,, $\Delta D$ is derived from the integral equation given by:
\begin{equation}\label{eq:DeltaD}
I = \int_{-\Delta D}^{\Delta D} \frac{1}{\sqrt{2\pi}\sigma} e^{-\frac{(x-\mu)^2}{2\sigma^2}} dx =90\%.
\end{equation}
%\FF{
The choice of $90\%$ confidence level lets us compare the obtained constraints on the exotic interactions with the previous results \cite{fadeev_pseudovector_2022, ficek_constraints_2018, karshenboim_hyperfine_2011}. However, following the suggestion made in the review \cite{cong_spin-dependent_2024}, in Supplemental Materials we also present results at the $95\%$-confidence level.
%}
%The $90\%$ confidence level \cite{ficek_constraints_2018,fadeev_pseudovector_2022} is chosen to give constraints for exotic interaction between neucleons in this work. %Note \cite{fadeev_pseudovector_2022} set constraints on the confidence level of 90\%. 

%Comparing $\Delta D$ from Ref.\,\cite{fadeev_pseudovector_2022} and the one from Table\,\ref{table1} yields a seven-fold improvement over the original constraints for $g_p^eg_p^p$ and $g_A^eg_A^p$ established by \citet{fadeev_pseudovector_2022} using $V_{AA}$ ($V_2+V_3$) and $V_{pp}$ ($V_3$). 

By comparing the value of $\Delta D$ from Ref.\,\cite{fadeev_pseudovector_2022} with our result in Table\,\ref{table1}, we find a seven-fold improvement in the constraints on the coupling constants $g_p^eg_p^p$ and $g_A^eg_A^p$, which were originally established by \citet{fadeev_pseudovector_2022} based on $V_{AA}$ (i.e., $V_2+V_3$) and $V_{pp}$ (i.e., $V_3$).

%\LC{
We further emphasize that the vector-vector interaction described by $(V_2+V_3)|_{VV}$ 
%$V_{VV}$ ($V_2+V_3$) 
%\FF{(Is there a difference between $V_{VV}$ ($V_2+V_3$) and $(V_2+V_3)|_{VV}$?} 
can and should also be explored within the framework.
%\FF{(I don't think a sentence should begin with the equation, we probably need to add something here.)}
This exotic potential has the form:
%}
\begin{widetext}
\begin{equation}
%\begin{aligned}
\label{gVgV_V23}
(V_2+V_3)|_{VV} =g_V^eg_V^p\frac{\hbar^3}{16\pi c m_em_p}\left[ \boldsymbol{\sigma}_e \cdot \boldsymbol\sigma_p \left[ \frac{1}{r^3} + \frac{1}{\lambda r^2} + \frac{1}{\lambda^2 r} - \frac{8 \pi}{3} \delta(\boldsymbol{r}) \right]  -  \left( \boldsymbol{\sigma}_e \cdot \hat{\boldsymbol{r}} \right) \left( \boldsymbol\sigma_p\cdot \hat{\boldsymbol{r}} \right)  \left( \frac{3}{r^3} + \frac{3}{\lambda r^2} + \frac{1}{\lambda^2 r} \right)  \right] e^{-{r}/{\lambda}} \, , 
%\end{aligned}
\end{equation}
\end{widetext}
%\LC{
where $\boldsymbol{\sigma}_{e}$ and $\boldsymbol{\sigma}_p$ are the Pauli vector operators corresponding to the spins $\boldsymbol{s}_i = \hbar \boldsymbol{\sigma}_i / 2$ of the electron and proton;
%$r$ denotes the distance between the two fermions;
$\mathbf{r}$ denoting the difference of positions of the fermions, $r=|\mathbf{r}|$ is the distance between them, while $\mathbf{\hat{r}}$ is the unit vector in the direction of $\mathbf{r}$;
$\lambda$ is the range of the force, given by $\lambda = \hbar / (M c)$, where $M$ is the mass of the new boson;
$m_e$ and $m_p$ represent the masses of the electron and proton, respectively.
%}
\begin{table}[!h]
\centering
\caption{Comparison of theoretical and experimental values of $D_{21}$ used to calculate $\Delta D$.
}\label{table1}
\begin{tabular}{lccc}
\hline
\hline
Parameter & $D_{21}$ &  $D_{21}$ \\
& Used in Ref.\,\cite{fadeev_pseudovector_2022} & Used in this work\\
\hline
Theory (kHz)& 48.953(3) \cite{karshenboim_improved_2005,karshenboim_study_2006} & 48.9541(23) \cite{yerokhin_electron_2008} \\
Expt (kHz)& 48.923(54) \cite{karshenboim_precision_2005,hellwig_measurement_1970} & 48.9592(68) \cite{bullis_ramsey_2023}  \\
\hline
$\mu$ (kHz)& 0.03 & -0.0051\\
$\sigma$ (kHz) & 0.054 & 0.0072\\
\hline
$\Delta D$ (kHz) (90\%) &0.102  & 0.0144\\
%$\Delta D$ (kHz) (90\%)& 0.105& 0.0195 \\
\hline
\hline
\end{tabular}
\end{table}

\begin{figure}[!htp]
\centering
\includegraphics[width=0.45\textwidth]{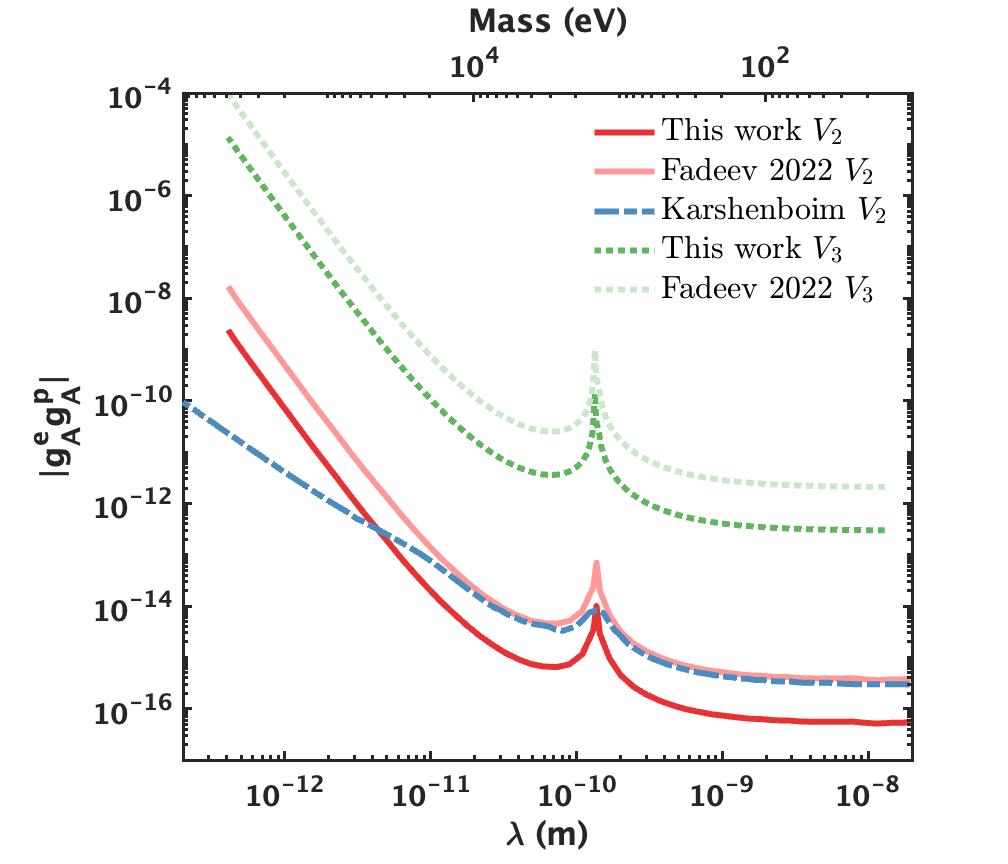}
\caption{Constraints on the coupling constants product $g^e_A g^p_A$.% as a function of the interaction range $\lambda$ shown on the bottom x-axis. The top x-axis represents the new spin-1 boson mass $M$. 
%\FF{
The results come from this work (dark solid red, dark dotted green curve for $V_2$ and $V_3$, respectively), \citet{fadeev_pseudovector_2022} (light solid red, light dotted green curve), and \citet{karshenboim_hyperfine_2011} (dashed-dotted blue curve).
%}
%Constraints for exotic interaction between $e$-$p$ are obtained from this work (dark solid red, dark dotted green curve) for $V_2$ and $V_3$, respectively, and \citet{fadeev_pseudovector_2022} (shallow solid red, shallow dotted green curve) and \citet{karshenboim_hyperfine_2011} (dashe-dotted blue curve), respectively.
}
\label{fig:gAgA}
\end{figure}
%\newpage
\begin{figure}%[!htp]
\centering
\includegraphics[width=0.45\textwidth]{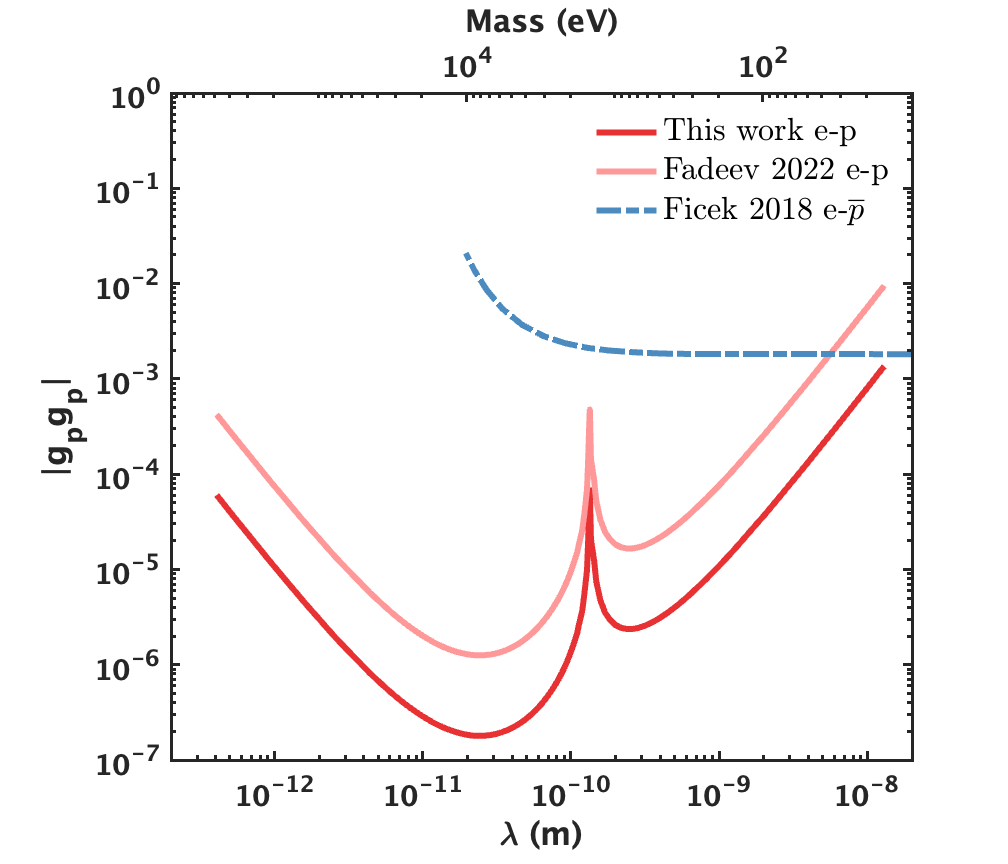}
\caption{
%\FF{
Constraints on the coupling constants products $g^e_p g^p_p$ and $g^e_p g^{\overline{p}}_p$ coming from the study of $V_3$.
% as a function of the interaction range $\lambda$ shown on the bottom x-axis. 
%The top x-axis represents the new spin-1 boson mass $M$.
The results for $e$-$p$ come from this work (dark solid red) and \citet{fadeev_pseudovector_2022} (shallow solid red), while the constraints for $e$-$\overline{p}$ are from \citet{ficek_constraints_2018} (dashed-dotted blue curve).
%}
%Constraints for exotic interaction between $e$-$p$ are obtained from this work (dark solid red) and \citet{fadeev_pseudovector_2022} (shallow solid red). The constraints for $e$-$\overline{p}$ from \citet{ficek_constraints_2018} (dashed-dotted blue curve) is presented for reference. Constraints for $g_pg_p$ are obtained by study $V_3$.
}
\label{fig:gpgp}
\end{figure}
\begin{figure}%[!htp]
\centering
\includegraphics[width=0.45\textwidth]{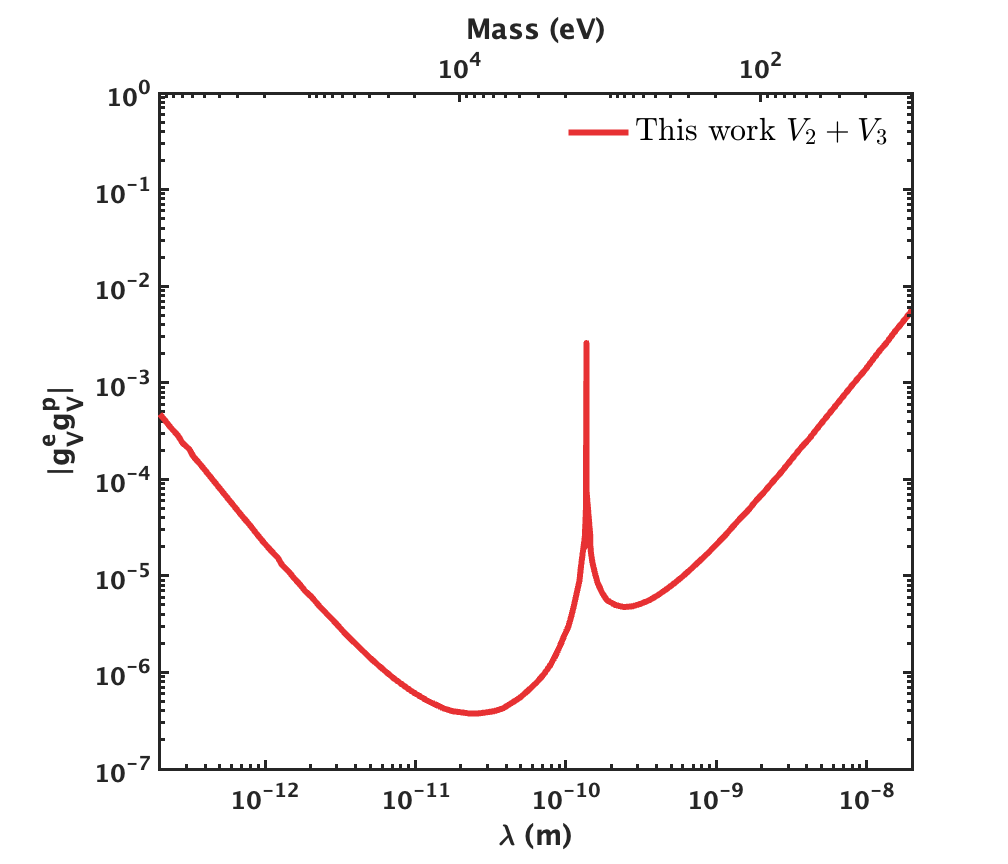}
\caption{Constraints on the coupling constants product $g^e_V g^p_V$.
% as a function of the interaction range $\lambda$ shown on the bottom x-axis.
%The top x-axis represents the new spin-1 boson mass $M$. %Constriants for exotic interaction between $e$-$p$ are obtained from this work (dark solid red) and \citet{fadeev_pseudovector_2022} (shallow solid red). The constraints for $e$-$\overline{p}$ from \citet{ficek_constraints_2017} (dashe-dotted blue curve) is presented for reference. Constraints for $g_Vg_V$ are obtained by study Eq.\,\eqref{gVgV_V23}.
}
\label{fig:gvgv}
\end{figure}

In Fig.\,\ref{fig:gAgA}-\ref{fig:gvgv}, we present our results, in comparison with those from Ref.\,\cite{fadeev_pseudovector_2022} as well as others \cite{karshenboim_hyperfine_2011,ficek_constraints_2018}.
%\FF{
They are presented as functions of the interaction range $\lambda$ (shown on the bottom x-axis with corresponding new spin-1 boson mass $M$ presented on the top x-axis) denoting the lower bound of the excluded region.
%}

For $g_A g_A$, Ref.\,\cite{fadeev_pseudovector_2022} originally obtained the constraints on $g_Ag_A$ by studying $V_2 + V_3$. Here, we redo the calculation and extract the constraints for $V_2$ and $V_3$ separately. \citet{karshenboim_hyperfine_2011} combined the constraints based on 1s hfs and $D_{21}$ from hydrogen and studied $V_2$, thus the $V_2$ result extracted  from Ref.\,\cite{fadeev_pseudovector_2022} is same as in \citet{karshenboim_hyperfine_2011} in the range of $\lambda> 10^{-11}$\,m. 
%\FF{
Additional constraints for $g_A^eg_A^p$ in the range $10^{3} \, \textrm{m}<\lambda< 10^{11}$\,m can be found in Ref.\,\cite{hunter_using_2013,hunter_using_2014}.
%}
%There are two other constraints for $g_A^eg_A^p$ from Ref.\,\cite{hunter_using_2013,hunter_using_2014} in the range $10^{3} \, \textrm{m}<\lambda< 10^{11}$\,m.

Regarding $g_p g_p$, Ref.\,\cite{fadeev_pseudovector_2022} previously provided the only constraints for exotic interactions between $e$-$p$. We present another result for $e$-$\overline{p}$ based on antiprotonic helium \cite{ficek_constraints_2018} for comparison.

%\LC{
For $g_V g_V$, we present the constraints obtained by studying Eq.\,\eqref{gVgV_V23}. As mentioned in our recent review \cite{cong_spin-dependent_2024}, while current constraints are approximately inferred from $V_3$, no existing result has addressed the combination $V_2 + V_3$, except our recent study on spin-spin coupling in tritium deuteride (DT) \cite{cong_constraints_2024}. We emphasize this point to draw the community’s attention to the importance of further studies focusing on $V_2 + V_3$ for vector-vector type exotic interactions.
%}
%\LC{One can also mention other types of potentials, but consider other two figures together.}

%\LC{
In summary, we have obtained improved constraints on electron–proton exotic interaction couplings $g_Ag_A$ and $g_pg_p$, and have set new constraints on $g_Vg_V$. Looking ahead, 
%} 
\citet{bullis_ramsey_2023} may achieve an additional tenfold improvement in precision by increasing the size of their setup \cite{rini_record_2023}, which would correspondingly further tighten the constraints on exotic electron-proton interactions. 

\section*{Acknowledgements}
This research was supported in part by the DFG Project ID 390831469: EXC 2118 (PRISMA+ Cluster of Excellence) and by the COST Action within the project COSMIC WISPers (Grant No. CA21106). 
F.\,F.~acknowledges the support of the Austrian Science Fund (FWF) via Project No.~P 36455 (DOI:10.55776/P36455) and Wittgenstein Award (DOI:10.55776/Z387)

\appendix
\section*{Supplemental Materials}
%\LC{
Note that we present the 90\% confidence level (CL) results in the figures of this work, as the other results used for comparison are given at the same level. However, following the suggestions made in the review \cite{cong_spin-dependent_2024}, we also calculated the 95\% CL results using Eq.\,\ref{eq:DeltaD}, obtaining $\Delta D = 0.0170$\,kHz. The 95\% CL results are available online \cite{cong_improved_2025} and are also included in the fifth force dataset \cite{cong_fifth_2024}.
\providecommand{\noopsort}[1]{}

\end{document}